\documentclass[preprint,12pt]{elsarticle} 
\usepackage{graphicx} 
\usepackage{amssymb} 
\usepackage{amsmath} 

\journal{Nuclear Physics A} 

\begin{document} 

\begin{frontmatter} 

\title{Three-body model calculations of $N\Delta$ and $\Delta\Delta$ 
dibaryon resonances} 
\author[a]{Avraham Gal} 
\author[b]{Humberto Garcilazo} 
\address[a]{Racah Institute of Physics, The Hebrew University, 91904 
Jerusalem, Israel} 
\address[b]{Escuela Superior de F\' \i sica y Matem\'aticas \\ 
Instituto Polit\'ecnico Nacional, Edificio 9, 07738 M\'exico D.F., Mexico} 

\begin{abstract} 

Three-body hadronic models with separable pairwise interactions are formulated 
and solved to calculate resonance masses and widths of $L=0$ $N\Delta$ and 
$\Delta\Delta$ dibaryons using relativistic kinematics. For $N\Delta$, 
$I(J^P)=1(2^+)$ and $2(1^+)$ resonances slightly below threshold are found 
by solving $\pi NN$ Faddeev equations. For $\Delta\Delta$, several resonances 
below threshold are found by solving $\pi N\Delta$ Faddeev equations in which 
the $N\Delta$ interaction is dominated by the $1(2^+)$ and $2(1^+)$ resonating 
channels. The lowest $\Delta\Delta$ dibaryon resonances found are for 
$I(J^P)=0(3^+)$ and $3(0^+)$, the former agreeing well both in mass and in 
width with the relatively narrow ${\cal D}_{03}(2370)$ resonance observed 
recently by the WASA@COSY Collaboration. Its spin-isospin symmetric partner 
${\cal D}_{30}$ is predicted with mass around 2.4 GeV and width about 80 MeV. 

\end{abstract}


\end{frontmatter} 

\section{Dedication} 
\label{sec:ded} 

This work is dedicated to the memory of Gerry Brown who has charted and 
shaped up the frontiers of Nuclear Physics for about half a century. 
Dibaryons, among many other topical subjects, fascinated Gerry and he 
has contributed imaginatively to this subject, too. We feel honored to 
add our modest contribution to this memorial issue of Nuclear Physics A. 

\newpage 

\section{Introduction} 
\label{sec:intro} 

Non-strange $s$-wave dibaryon resonances ${\cal D}_{IS}$ with isospin $I$ and 
spin $S$ were predicted by Dyson and Xuong \cite{dyson64} in 1964 as early 
as SU(6) symmetry for baryons, placing the nucleon $N(939)$ and its $P_{33}$ 
$\pi N$ resonance $\Delta(1232)$ in the same ${\bf 56}$ multiplet, proved 
successful. These authors chose the ${\bf 490}$ lowest-dimension SU(6) 
multiplet in the $\bf{56\times 56}$ direct product containing the SU(3)-flavor 
$\overline{\bf 10}$ and ${\bf 27}$ multiplets in which the deuteron 
${\cal D}_{01}$ and $NN$ virtual state ${\cal D}_{10}$ are classified. 
This gave four non-strange dibaryon candidates with masses listed in 
Table~\ref{tab:dyson} in terms of constants $A,B$. Identifying $A$ with the 
$NN$ threshold mass 1878~MeV, the value $B\approx 47$~MeV was derived by 
assigning ${\cal D}_{12}$ to the $pp\leftrightarrow \pi^+ d$ coupled-channel 
resonance behavior noted then at 2160~MeV, near the $N\Delta$ threshold 
(nominally 2.171~MeV). This led in particular to a predicted mass $M=2350$~MeV 
for ${\cal D}_{03}$, followed since 1977 by many quark-based model 
calculations as reviewed by us recently \cite{gg14}. 

\begin{table}[hbt]
\begin{center}
\caption{SU(6)-predicted masses of non-strange $L=0$ dibaryons ${\cal D}_{IS}$ 
with isospin $I$ and spin $S$, using the Dyson-Xuong mass formula 
$M=A+B[I(I+1)+S(S+1)-2]$ \cite{dyson64}.}
\begin{tabular}{ccccccc}
\hline
${\cal D}_{IS}$ & ${\cal D}_{01}$ & ${\cal D}_{10}$ & ${\cal D}_{12}$ &
${\cal D}_{21}$ & ${\cal D}_{03}$ & ${\cal D}_{30}$ \\
\hline
$BB'$ &$NN$&$NN$& $N\Delta$ & $N\Delta$ & $\Delta\Delta$ & $\Delta\Delta$ \\
SU(3)$_{\rm f}$ & $\overline{\bf 10}$ & ${\bf 27}$ & ${\bf 27}$ & ${\bf 35}$ &
$\overline{\bf 10}$ & ${\bf 28}$ \\
$M({\cal D}_{IS})$ & $A$ & $A$ & $A+6B$ & $A+6B$ & $A+10B$ & $A+10B$ \\
\hline
\end{tabular}
\label{tab:dyson}
\end{center}
\end{table}

The ${\cal D}_{12}$ dibaryon conjectured by Dyson and Xuong \cite{dyson64} 
shows up in the $^1D_2$ nucleon-nucleon partial wave above the $\pi NN$ 
threshold and it is produced by the coupling between the $d$-wave $NN$ channel 
and the $s$-wave $N\Delta$ channel where $\Delta$ is the pion-nucleon $P_{33}$ 
resonance, i.e. the coupling between the two-body $NN$ channel and the 
three-body $\pi NN$ channel. Representative values (in MeV) derived 
phenomenologically in Refs.~\cite{igor84,arndt87,hosh92} for the pole position 
$W=M-{\rm i}\Gamma /2$ of ${\cal D}_{12}$ are 
\begin{equation} 
(M, \Gamma): \;\;\;\;\; (2176\pm 6,107\pm 23), \;\;\;\;(2148,126), \;\;\;\; 
(2144,110), 
\label{eq:D12} 
\end{equation} 
respectively, in good agreement with the mass value used in 
Ref.~\cite{dyson64}. 
Another positive-parity dibaryon, with quantum numbers $IJ=03$, has been 
observed at $\sqrt{s}=2.37$~GeV in a kinematically complete measurement of 
the pion-production reaction $np\to d\pi^0\pi^0$ \cite{wasa11}. Viewed as the 
$\Delta\Delta$ dibaryon quasibound state ${\cal D}_{03}$ it is deeply bound, 
by about 90 MeV with respect to the $\Delta\Delta$ threshold. An equally 
intriguing feature of this dibaryon resonance is its relatively small width 
$\Gamma({\cal D}_{03})\approx 70$~MeV, considerably below the phase-space 
expectation $\Gamma_{\Delta}\leq\Gamma({\cal D}_{03})\leq 2\Gamma_{\Delta}$, 
with $\Gamma_{\Delta}\approx 120$~MeV. The binding energy of ${\cal D}_{03}$ 
has been calculated in several works using various one-boson-exchange 
potential (OBEP) models \cite{kamae77,ueda78,sato83} and a variety of 
quark-based models for the (real) $\Delta\Delta$ interaction \cite
{oka80,mulders80,mulders83,malt85,gold89,valc01,mota02,ping02,ping09,huang13} 
leading to binding energies running from a few MeV up to several hundred MeV. 
However, no calculation other than the one reported by us recently \cite{gg13} 
has so far been able to explain its small width. 

In the present paper we extend the hadronic model constructed by us for the 
$\Delta\Delta$ dibaryon resonance ${\cal D}_{03}$ \cite{gg13} in order to 
study systematically all the $s$-wave $N\Delta$ and $\Delta\Delta$ dibaryon 
candidates. With isospin $\frac{1}{2}$ and spin $\frac{1}{2}$ for nucleons, 
and isospin $\frac{3}{2}$ and spin $\frac{3}{2}$ for $\Delta$'s, the allowed 
range of isospin $I$ and total angular momentum $J=S$ values consists of 
$IJ=12,21,11,22$ for $N\Delta$, and $IJ=01,03,10,12,21,23,30,32$ for 
$\Delta\Delta$ in consequence of the Pauli principle requirement $I+J=$~odd 
for two identical $\Delta$'s.  

Considering the $\Delta$ as a $\pi N$ resonance, it is straightforward to 
replace the $\Delta N$ system by a $\pi N N$ system of three stable particles 
for which Faddeev equations with separable pairwise potentials may be applied 
to calculate the mass and width of the various $N\Delta$ resonance candidates 
enumerated above. This program is followed in Sect.~\ref{sec:NDel}. 
For the $\Delta\Delta$ system, if we wish to keep applying three-body Faddeev 
equations rather than resorting to the more complicated $\pi N\pi N$ four-body 
Faddeev-Yakubovsky equations, it is necessary to treat initially one of the 
$\pi N$ pairs by a stable $\Delta$ within a $\pi N\Delta$ three-body model, 
recovering its decay-width contribution in the last stage of the dibaryon mass 
and width calculation. This program is followed in Sect.~\ref{sec:DelDel}. 
Finally, in Sect.~\ref{sec:concl} we summarize our work and present 
additional discussion.

\section{$N\Delta$ dibaryons} 
\label{sec:NDel} 

The $N\Delta$ system in which $N$ and $\Delta$ are in a relative orbital 
angular momentum state $\lambda=0$ is a three-body system consisting of 
a pion and two nucleons, where the $\pi N$ subsystem is dominated by the 
$P_{33}$ resonant channel (the $\Delta$ resonance) and the $NN$ subsystem 
is dominated by the $^3S_1$ and $^1S_0$ channels. We work in momentum space 
using Jacobi vector coordinates ${\vec p}_k, {\vec q}_k$ to denote the 
relative momentum of pair $(i,j)$ and that of particle $k$ with respect to 
the center of mass (cm) of pair $(i,j)$, respectively, with $(i,j,k)$ cyclic 
permutation of (1,2,3). Thus, labeling the pion as particle 1 and the two 
nucleons as particles 2 and 3, ${\vec p}_1$ is the $NN$ relative momentum and 
${\vec q}_1$ is the pion momentum with respect to the cm of the $NN$ pair. 

\subsection{Two-body interactions} 

We use separable pairwise interactions fitted to phase shifts in the dominant 
channels, as deduced from elastic scattering data. Thus, the $\pi N$ 
interaction which is dominated by the $P_{33}$ channel at relevant energies 
is represented by a rank-one separable potential   
\begin{equation} 
V_3(p_3,p'_3)=\lambda_3 g_3(p_3)g_3(p_3^\prime), 
\label{eq6} 
\end{equation} 
so that solving the Lippmann-Schwinger equation with relativistic 
kinematics one obtains a similar form: 
\begin{equation} 
t_3(\omega_3;p_3,p_3^\prime)= g_3(p_3)\tau_3(\omega_3)g_3(p_3^\prime), 
\label{eq8} 
\end{equation} 
with 
\begin{equation} 
\tau_3^{-1}(\omega_3)=\lambda_3^{-1}-\int_0^\infty p_3^2 dp_3 
\frac{[g_3(p_3)]^2} 
{\omega_3-E_N(p_3)-E_{\pi}(p_3)+i\epsilon},  
\label{eq9} 
\end{equation} 
where $E_h(p)=\sqrt{m_h^2+p^2}$ for hadron $h$ with mass $m_h$. Here, 
$\tau_3(\omega_3)$ is the propagator of the $\Delta$ isobar in the 
pion-nucleon cm system, with $\omega_3$ the two-body $\pi N$ cm energy. 
In the three-body cm system, with $W$ the total three-body cm energy and 
$q_3$ the momentum of the spectator nucleon with respect to the two-body 
$\pi N$ isobar, this propagator becomes a function of both $W$ and $q_3$ 
and its inverse is given by 
\begin{equation} 
{\cal T}_3^{-1}(W;q_3)=\lambda_3^{-1}-\int_0^\infty p_3^2 dp_3 
\frac{[g_3(p_3)]^2} 
{W-{\cal E}_3(p_3,q_3)-E_N(q_3)+i\epsilon}, 
\label{eq9ppp} 
\end{equation} 
where ${\cal E}_3(p_3,q_3)=\sqrt{[E_{\pi}(p_3)+E_N(p_3)]^2+q_3^2}$. 
For $q_3=0$, when the three-body cm system degenerates to the two-body cm 
system, ${\cal T}_3$ reduces to $\tau_3$ with a shifted value of energy: 
${\cal T}_3(W;q_3=0)=\tau_3(W-m_N)$. 

We considered two different parametrizations for the form factor $g_3$. 
Type I is defined by 
\begin{equation} 
g_3(p_3)=p_3{\rm exp}(-p_3^2/\beta_3^2) 
           +A_3p_3^3{\rm exp}(-p_3^2/\alpha_3^2). 
\label{eq7} 
\end{equation} 
This form factor falls off exponentially upon $p_3\to\infty$. 
Type II is defined by 
\begin{equation} 
g_3(p_3)=\frac{p_3}{(1+p_3^2/\beta_3^2)^2} 
           +A_3\frac{p_3^3}{(1+p_3^2/\alpha_3^2)^3}, 
\label{eq7p} 
\end{equation} 
which falls off as $p_3^{-3}$ upon $p_3\to\infty$. The parameters of these 
two models were fitted to the $\pi N$ $P_{33}$ phase shifts from Arndt et al. 
\cite{arndt06} and are listed in Table~\ref{tab:piN}. The fit of Type I is 
shown in Fig.~2 of Ref.~\cite{gg11} and the fit of Type II looks essentially 
identical to that of type I. The table also lists the distance $r_0$ at 
which the Fourier transform ${\tilde g}_3(r)$ flips sign, which roughly 
represents the spatial extension of the $P_{33}$ $p$-wave form factor as 
discussed in Ref.~\cite{gg11}. 

\begin{table}[htb]  
\begin{center} 
\caption{Separable-potential parameters of the $\pi N$ $P_{33}$ form 
factor $g_3(p)$ (\ref{eq6}) fitted to phase shifts \cite{arndt06}, 
and the zero $r_0$ of the Fourier transform ${\tilde g}_3(r)$ \cite{gg11}, 
for two types of $g_3(p)$ labeled I (\ref{eq7}) and II (\ref{eq7p}).} 
\begin{tabular}{cccccc} 
\hline 
 type & $\lambda_3\;({\rm fm}^4)$ & $\beta_3\;({\rm fm}^{-1})$ &  
$\alpha_3\;({\rm fm}^{-1})$  & $A_3\;({\rm fm}^2)$ & $r_0\;({\rm fm})$  \\ 
\hline 
 I  & $-$0.07587 & 1.04 & 2.367 & 0.23 & 1.36      \\ 
 II & $-$0.04177 & 1.46 & 4.102 & 0.11 & 0.91    \\ 
\hline 
\end{tabular} 
\label{tab:piN} 
\end{center} 
\end{table} 

\begin{figure}[hbt] 
\begin{center} 
\includegraphics[width=0.48\textwidth]{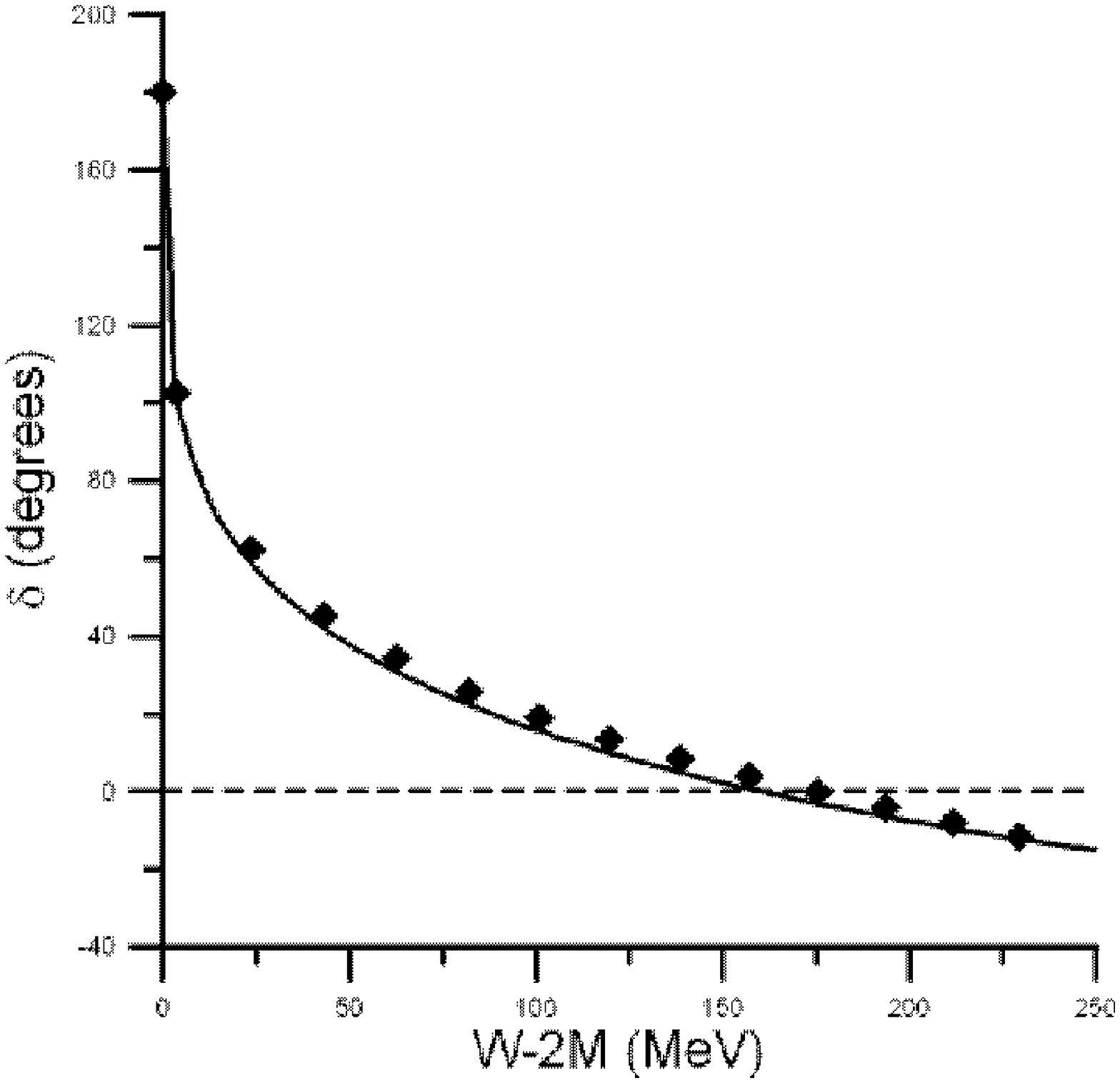} 
\includegraphics[width=0.48\textwidth]{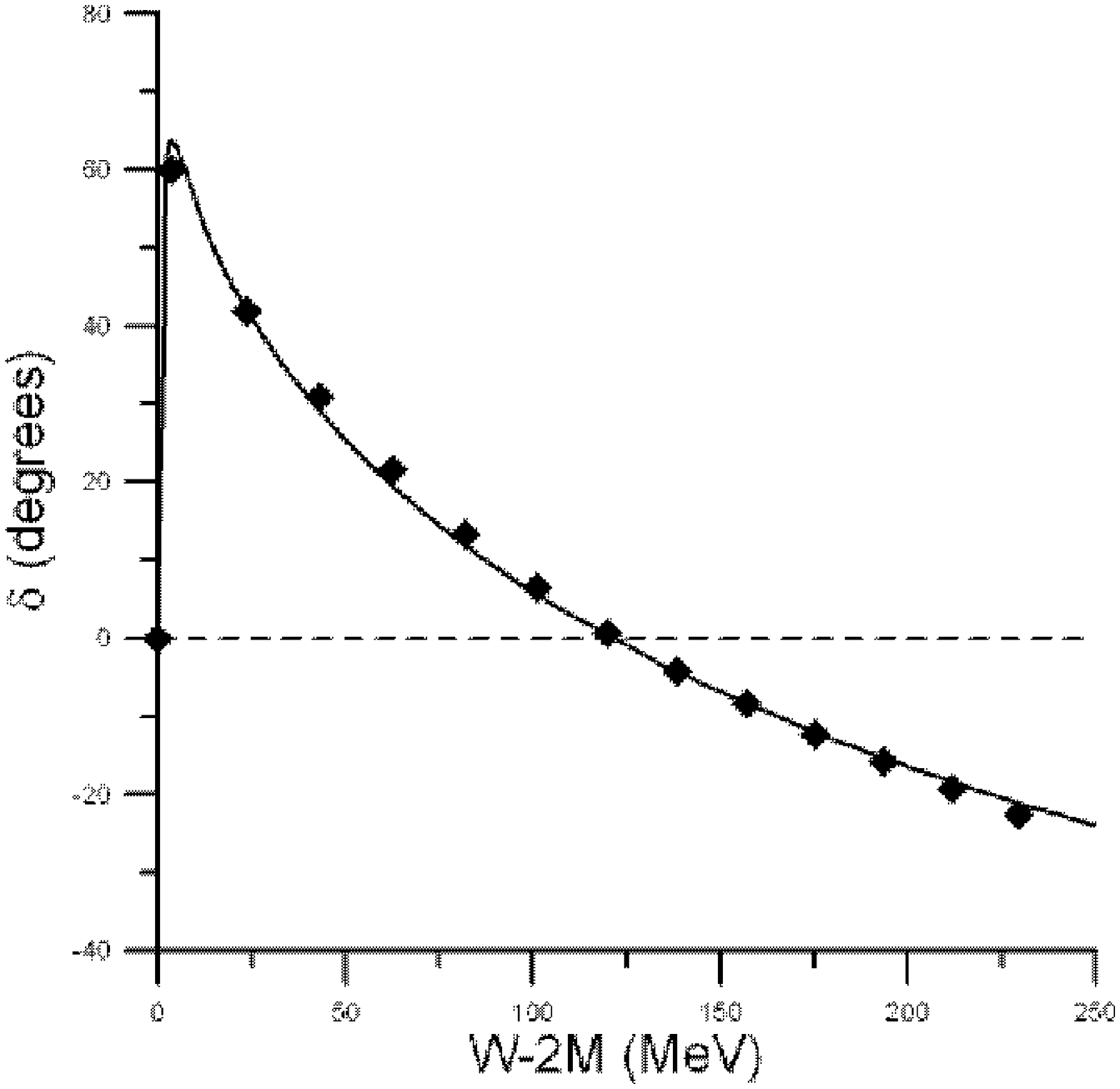} 
\caption{Fits of $NN$ separable potentials to $^3S_1$ (left) and $^1S_0$ 
(right) phase shifts \cite{arndt07}.} 
\label{fig:NN} 
\end{center} 
\end{figure}

For the $NN$ interaction we used rank-two separable potentials consisting 
of one attractive term and one repulsive term: 
\begin{equation} 
V_1^{\gamma}(p_1,p'_1)=\sum_{m=1}^2 \lambda_{1\gamma}^m g_{1\gamma}^m(p_1) 
g_{1\gamma}^m(p'_1), 
\label{eq15} 
\end{equation} 
where $\lambda_{1\gamma}^1$ is negative and $\lambda_{1\gamma}^2$ is 
positive in both fits of the $^3S_1$ ($\gamma=1$) and $^1S_0$ ($\gamma=2$) 
phase shifts. The resulting $t$ matrix is also separable, as follows: 
\begin{equation} 
t_1^{\gamma}(\omega;p_1,p'_1)=\sum_{m,n=1}^2 g_{1\gamma}^m(p_1) 
\tau_{1\gamma}^{mn}(\omega)g_{1\gamma}^n(p'_1), 
\label{eq16} 
\end{equation} 
\begin{equation} 
\tau_{1\gamma}^{mn}(\omega)=\frac{G_{1\gamma}^{3-m,3-n}(\omega)}
{G_{1\gamma}^{11}(\omega)G_{1\gamma}^{22}(\omega)-G_{1\gamma}^{12}(\omega)
G_{1\gamma}^{21}(\omega)}, 
\label{eq17} 
\end{equation} 
\begin{equation} 
G_{1\gamma}^{mn}(\omega)=\frac{1}{\lambda_{1\gamma}^{m}}\delta_{mn} 
-(-)^{m+n} \int_0^\infty p_1^2 dp_1 
\frac{g_{1\gamma}^m(p_1)g_{1\gamma}^n(p_1)}{\omega-2E_N(p_1)+i\epsilon}. 
\label{eq18} 
\end{equation} 
Form factors of the Yamaguchi type 
\begin{equation} 
g_{1\gamma}^m(p_1)=\frac{1}{p_1^2+(\alpha_{1\gamma}^m)^2} 
\label{eq18p} 
\end{equation} 
were fitted to the deuteron binding energy and the nucleon-nucleon $^3S_1$ 
and $^1S_0$ phase shifts. The parameters of these $NN$ potentials are 
given in Table~\ref{tab:NN} and the calculated phase shifts are compared 
in Fig.~\ref{fig:NN} with those deduced from experiment by Arndt 
et al.~\cite{arndt07}. 

\begin{table}[htb] 
\begin{center} 
\caption{Parameters of the rank-two nucleon-nucleon separable 
potential in the $^3S_1$ and $^1S_0$ partial waves fitted to 
$NN$ phase shifts \cite{arndt07}.} 
\begin{tabular}{ccccccc} 
\hline 
& channel ($\gamma$) & $\lambda_{1\gamma}^1\;({\rm fm}^{-2})$ & 
$\lambda_{1\gamma}^2\;({\rm fm}^{-2})$ & $\alpha_{1\gamma}^1\;({\rm fm}^{-1})$ 
& $\alpha_{1\gamma}^2\;({\rm fm}^{-1})$ & \\ 
\hline 
& $^3S_1$ ($\gamma=1$) & $-$5.6  & 196.75 & 1.88 & 5.38  &    \\ 
& $^1S_0$ ($\gamma=2$) & $-$6.0  & 12411 & 1.90 & 5.60  &    \\ 
\hline 
\end{tabular} 
\label{tab:NN} 
\end{center} 
\end{table} 

\subsection{Faddeev equations of the $\pi NN$ system} 

\begin{figure}[hbt] 
\begin{center} 
\includegraphics[width=\textwidth]{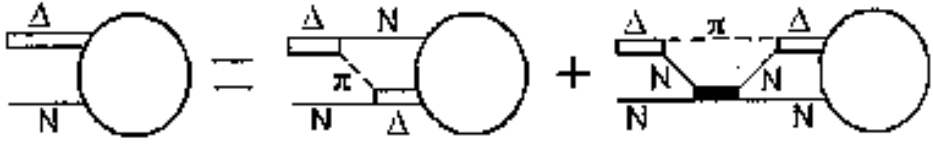} 
\caption{Diagrammatic representation of the $\pi NN$ Faddeev equations solved 
in the present work to calculate $N\Delta$ dibaryon resonance poles.} 
\label{fig:piNN} 
\end{center} 
\end{figure} 

In the case of the $\pi NN$ system with separable pairwise potentials, since 
two of the constituents are identical fermions, the Faddeev integral equations 
reduce to a single integral equation for the $N\Delta$(isobar) $T$ matrix 
shown diagrammatically in Fig.~\ref{fig:piNN}. For a positive-parity $\pi NN$ 
state with total isospin $I$ and angular momentum $J$, this equation is 
written explicitly as \cite{gg11} 
\begin{eqnarray} 
T^{IJ}(W;q_3)&=&\int_0^\infty dq_3^\prime M^{IJ}(W;q_3,q_3^\prime)
{\cal T}_3(W;q_3^\prime)T^{IJ}(W;q_3^\prime), \label{eq19} \\ 
M^{IJ}(W;q_3,q_3^\prime)&=&K_{23}^{IJ}(W;q_3,q_3^\prime)
+2\sum_{mn\gamma}\int_0^\infty K_{31;m\gamma}^{IJ}(W;q_3,q_1) \nonumber \\ 
&& \times {\cal T}_{1\gamma}^{mn}(W;q_1)K_{13;n\gamma}^{IJ}(W;q_1,q_3^\prime) 
dq_1, 
\label{eq20} 
\end{eqnarray} 
with a kernel $M^{IJ}$ given in terms of one-particle-exchange 
amplitudes $K_{ij}$: 
\begin{eqnarray} 
K_{23}^{IJ}(W;q_3,q'_3)&=& 
\frac{1}{2}q_3q'_3\int_{-1}^1 d{\rm cos}\theta\, 
g_3(p_3)\, g_3(p'_3)b_{23}^{IJ} 
\nonumber \\ && \times 
\frac{\hat p_3\cdot\hat p'_3}{W-E_N(q_3) 
-E_{\pi}(\vec q_3+{\vec q_3}^{\;\prime}) 
-E_N(q'_3)+i\epsilon}, 
\label{eq21} 
\end{eqnarray} 
\begin{eqnarray} 
K_{31;m\gamma}^{IJ}(W;q_3,q_1)&=& 
 \frac{1}{2}q_3q_1\int_{-1}^1 d{\rm cos}\theta\, 
g_3(p_3)\, g_{1\gamma}^m(p_1)b_{31;\gamma}^{IJ} 
\nonumber \\ && \times 
\frac{\hat p_3\cdot\hat q_1}{W-E_N(q_3)
-E_N(\vec q_1+\vec q_3)
-E_{\pi}(q_1)+i\epsilon},  
\label{eq22} 
\end{eqnarray} 
with $K_{13;m\gamma}^{IJ}(W;q_1,q_3)=K_{31;m\gamma}^{IJ}(W;q_3,q_1)$. 
The momenta $\vec p_3, {\vec p_3}^{\;\prime}$ in Eq.~(\ref{eq21}) and 
$\vec p_3, \vec p_1$ in Eq.~(\ref{eq22}) are the pairs relative momenta, 
given for relativistic kinematics in terms of $q_i, q_j$ and $\cos\theta$ 
by Eqs.~(39)--(43) in Ref.~\cite{gg11}. 
The factor 2 in Eq.~(\ref{eq20}) counts the two nucleons, each of which 
can be exchanged. In Eq.~(\ref{eq21}), $\theta$ is the angle between 
$\vec q_3$ and ${\vec q_3}^{\;\prime}$, whereas in Eq.~(\ref{eq22}) it is the 
angle between ${\vec q}_1$ and ${\vec q}_3$. Finally, the isospin and 
angular-momentum recoupling coefficients $b_{ij}^{IJ}$ in Eqs.~(\ref{eq21}) 
and (\ref{eq22}) are given by \cite{gar06} 
\begin{eqnarray} 
b_{ij}^{IJ} &=& 
(-)^{I_{ik}+I_j-I}\sqrt{(2I_{ik}+1)(2I_{jk}+1)} \, 
W(I_j I_k I I_i ; I_{jk} I_{ik}) \nonumber \\ && \times 
(-)^{J_{ik}+J_j-J}\sqrt{(2J_{ik}+1)(2J_{jk}+1)} \, 
W(J_j J_k J J_i ; J_{jk} J_{ik}), 
\label{e8c5} 
\end{eqnarray} 
where $W$'s are Racah coefficients in terms of isospins $I_1=1,I_2=I_3=
\frac{1}{2}$ with $I_{12}=I_{13}=\frac{3}{2}$ and, independently, angular 
momenta $J_1=1,J_2=J_3=\frac{1}{2}$ with $J_{12}=J_{13}=\frac{3}{2}$. 
Note, however, that $I_{23}=0$ is correlated with $J_{23}=1$ ($\gamma=1$) 
and $I_{23}=1$ with $J_{23}=0$ ($\gamma=2$). The suffix $\gamma$ in 
$b_{31;\gamma}^{IJ}$ keeps track of this correlation. These coefficients are 
listed in Table~\ref{tab:bij}. 

\begin{table}[hbt] 
\begin{center} 
\caption{Recoupling coefficients $b_{ij}^{IJ}$ (\ref{e8c5}) for $\pi NN$ 
Faddeev calculations. The value listed for $b_{31}^{11}$ is independent of 
the suffix $\gamma$, see text.} 
\begin{tabular}{cccccc} 
\hline 
$b_{23}^{12}$ & $b_{31}^{12}$ & $b_{23}^{21}$ & $b_{31}^{21}$ & 
$b_{23}^{11}$ & $b_{31}^{11}$ \\ 
\hline 
$-$1/3 & $\sqrt{2/3}$ & $-$1/3 & $\sqrt{2/3}$ & 1/9 & $-\sqrt{2/9}$ \\ 
\hline 
\end{tabular} 
\label{tab:bij} 
\end{center} 
\end{table} 

\subsection{Results and Discussion} 

In order to search for $\pi NN$ resonances, the integral equation 
(\ref{eq19}) was extended into the complex plane, using the standard 
procedure $q_i\to q_i\exp(-{\rm i}\phi)$ \cite{afnan84} which opens large 
sections of the unphysical sheet so that one can search for eigenvalues 
of the form $W=M-{\rm i}\frac{\Gamma}{2}$. 

Of the four possible $N\Delta$ $s$-wave states with $IJ=$ $12$, $21$, $11$, 
$22$, the last two are found not to resonate. This is easy to understand for 
the $IJ=$ $22$ state which cannot benefit from the $s$-wave $NN$ interactions 
in the $^3S_1$ and $^1S_0$ channels. In the case of the $IJ=$ $11$ state, 
since $b_{23}^{11}=\frac{1}{9}$ (see Table~\ref{tab:bij}), the $K_{23}^{11}$ 
amplitude (\ref{eq21}) is {\it repulsive}, and with $(b_{31}^{11})^2=
\frac{2}{9}$ the other component of the kernel $M^{11}$ (\ref{eq20}) is too 
weak to provide sufficient attraction to generate resonances. 

The $N\Delta$ states with $IJ=$ $12$ and $21$ are found to resonate. 
We note that only $^3S_1$ enters the calculation of the $IJ=$ $12$ resonance, 
while for the $21$ resonance calculation only $^1S_0$ enters. Furthermore, 
$b_{23}^{12}=b_{23}^{21}$ and $b_{31}^{12}=b_{31}^{21}$, so that if the 
$^3S_1$ and $^1S_0$ interactions were equal, the $IJ=12$ and $IJ=21$ 
resonances would have been degenerate. However, since the $^3S_1$ interaction 
is more attractive than the $^1S_0$ interaction, one expects that the $IJ=12$ 
resonance lies below the $IJ=21$ resonance. For the $P_{33}$ interaction 
model of type I (\ref{eq7}), the $IJ=12$ resonance indeed lies 18~MeV below 
the $IJ=21$ resonance, whereas for type II $P_{33}$ interaction model 
(\ref{eq7p}), the difference shrinks to merely 10~MeV, as inferred from the 
calculated masses listed in Table~\ref{tab:NDel}. These listed mass values 
for $IJ=12$ and $IJ=21$ are sufficiently close to each other to qualify as 
approximately degenerate. 

\begin{table}[hbt] 
\begin{center} 
\caption{$N\Delta$ dibaryon $S$-matrix pole position 
$W=M-{\rm i}\frac{\Gamma}{2}$ (in MeV) for ${\cal D}_{12}$ and 
${\cal D}_{21}$, obtained by solving $\pi NN$ Faddeev equations 
for two choices of the $\pi N$ $P_{33}$ form factor, type I 
(\ref{eq7}) and type II (\ref{eq7p}) marked by superscripts. 
The last column lists the results of a nonrelativistic Faddeev 
calculation by Ueda \cite{ueda82}.} 
\begin{tabular}{ccccccc} 
\hline 
$W^{\rm I}({\cal D}_{12})$ & $W^{\rm I}({\cal D}_{21})$ &  &
$W^{\rm II}({\cal D}_{12})$ & $W^{\rm II}({\cal D}_{21})$ &  & 
$W^{\rm Ueda}({\cal D}_{12})$  \\  
\hline 
2147$-{\rm i}$60 & 2165$-{\rm i}$64 &  &  2159$-{\rm i}$70 & 
2169$-{\rm i}$69 &  & 2116$-{\rm i}$61  \\
\hline 
\end{tabular} 
\label{tab:NDel} 
\end{center} 
\end{table} 

We note that the calculated half-widths listed in the table are close to the 
half-width of the free $\Delta$, as expected naively from a loosely bound 
$N\Delta$ system. This is also expected within a $\pi NN$ model provided the 
$\pi N$ spatial extension is sufficiently small compared to the $NN$ average 
distance. If the pion's wavelength were commensurate with the $NN$ average 
distance, the decay width of the $\pi NN$ system would have exceeded the 
free $\Delta$'s width, up to ideally twice as much. 

The mass and width values calculated for the $IJ=12$ resonance 
lie comfortably within the range of values exhibited in Eq.~(\ref{eq:D12}) 
for the phenomenologically deduced ${\cal D}_{12}$ dibaryon. For this 
reason we associate the $IJ=12$ and $IJ=21$ $\pi NN$ poles found here 
with the ${\cal D}_{12}$ and ${\cal D}_{21}$ dibaryon candidates of 
Table~\ref{tab:dyson} and Eq.~(\ref{eq:D12}). Finally, in the last column 
of the table we list the result of a $\pi NN$ Faddeev calculation for 
${\cal D}_{12}$ by Ueda \cite{ueda82} using nonrelativistic kinematics. 
Ueda's calculated mass comes about 30 to 40 MeV below the values calculated 
by us, in rough agreement with our own experience in comparing Faddeev 
calculations that use relativistic kinematics to similar ones using 
nonrelativistic kinematics \cite{gg12}.

\section{$\Delta\Delta$ dibaryons} 
\label{sec:DelDel}

Our main interest in this section is in $\Delta\Delta$ dibaryon candidates, 
particularly the ${\cal D}_{03}$ and ${\cal D}_{30}$ predicted by Dyson and 
Xuong \cite{dyson64}, see Table~\ref{tab:dyson}. As shown in the previous 
section, describing $N\Delta$ systems in terms of a stable nucleon ($N$) and 
a two-body $\pi N$ resonance ($\Delta$) leads to a well defined $\pi NN$ 
three-body model in which $IJ=12$ and $21$ resonances are generated. These 
were identified by us with the ${\cal D}_{12}$ and ${\cal D}_{21}$ dibaryons 
of Table~\ref{tab:dyson} and Eq.~(\ref{eq:D12}). This relationship between 
$N\Delta$ and $\pi NN$ may be generalized into relationship between a two-body 
$B\Delta$ system and a three-body $\pi NB$ system, where the baryon $B$ stands 
for $N, \Delta, Y$ (hyperon) etc. In order to stay within a three-body 
formulation we need to assume that the baryon $B$ is stable. For $B=N$, this 
formulation reduces to the one discussed in the previous section for $N\Delta$ 
dibaryons. For $B=\Delta$, once properly formulated, it relates the 
$\Delta\Delta$ system to the three-body $\pi N\Delta$ system, suggesting to 
seek $\Delta\Delta$ dibaryon resonances by solving $\pi N\Delta$ Faddeev 
equations, with a stable $\Delta$. The decay width of the $\Delta$ resonance 
will have to be considered at the penultimate stage of the calculation. 
In terms of two-body isobars we have then a coupled-channel problem 
\begin{equation} 
B\Delta\leftrightarrow\pi D, 
\label{eq:D} 
\end{equation} 
where $D$ stands generically for appropriate dibaryon isobars: ${\cal D}_{01}$ 
and ${\cal D}_{10}$, which are the $NN$ isobars identified with the deuteron 
and virtual state respectively, for $B=N$; ${\cal D}_{12}$ and ${\cal D}_{21}$ 
for $B=\Delta$.  

\begin{figure}[hbt] 
\begin{center} 
\includegraphics[width=0.7\textwidth]{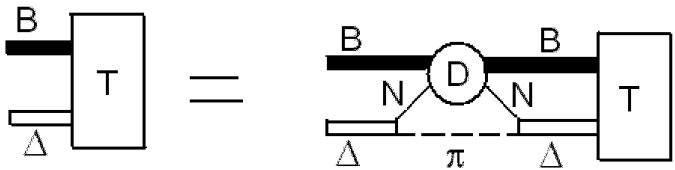} 
\caption{Diagrammatic representation of the $\pi NB$ Faddeev equations solved 
to calculate $B\Delta$ dibaryon resonance poles.} 
\label{fig:piNB} 
\end{center} 
\end{figure} 

Within the set of Faddeev equations for three {\it stable} particles $\pi$, 
$N$ and $B$, we label the $\pi$ meson as particle 1, the nucleon $N$ as 
particle 2 and the stable baryon $B$ as particle 3, and let these particles 
interact pairwise through separable potentials. The interaction $V_3$ between 
$\pi$ and $N$ is limited to the $P_{33}$ channel which is dominated by the 
$\Delta$ resonance. Similarly, the interaction $V_1$ between $N$ and $B$, 
for $B=\Delta$, is limited to the $IJ=12$, $21$ channels corresponding to 
the ${\cal D}_{12}$ and ${\cal D}_{21}$ dibaryon resonances calculated in the 
previous section. Finally, the interaction $V_2$ between the $\pi$ meson and 
$B$ is neglected for $B=\Delta$, for lack of known isobar resonances 
to dominate it. Within this model, the coupled-channel $B\Delta -\pi D$ 
eigenvalue problem reduces, again, to a single integral equation for the 
$B\Delta$ $T$ matrix shown diagrammatically in Fig.~\ref{fig:piNB}, where 
starting with a $B\Delta$ configuration the $\Delta$-resonance isobar decays 
into $\pi N$, followed by $NB\to NB$ scattering through the $D$-isobar with a 
spectator pion, and ultimately by means of the inverse decay $\pi N\to\Delta$ 
back into the $B\Delta$ configuration. 

Since ${\cal D}_{12}$ in the $IJ=12$ channel appears as a resonance in the 
$NN$ $^1D_2$ partial wave, we will adjust the $NB$ separable potential to 
that piece of experimental information. In the case of the $IJ=21$ channel, 
unfortunately, there is no corresponding experimental information available 
so that we will have to rely on theoretical arguments based on the similarity 
between the channels $IJ=12$ and $IJ=21$. 

\subsection{Quantum statistics correlations} 

The formulation of Faddeev equations for the $\pi NB$ system requires that 
$B$ is a stable particle. For $B=\Delta$ we would like to grant $\Delta$ 
a {\it complex} mass, given by its $S$-matrix pole position, when appearing 
as spectator in the $\pi N$ propagator. By doing so we hope to provide a more 
realistic estimate of the decay width of $\Delta\Delta$ dibaryons. The width 
contribution of one of the $\Delta$ resonances is fully accounted for by the 
$\pi N$ isobar that represents it in the three-body model. Care must be 
exercised, however, to impose the necessary quantum statistics correlations 
between this pre-existing $N\pi$ pair and the $N\pi$ pair resulting from the 
other $\Delta$ decay. For ${\cal D}_{03}$, for example, assuming $s$-wave 
nucleons and $p$-wave pions implies space-spin symmetry for nucleons as well 
as for pions. With total $I$=0, Fermi-Dirac (Bose-Einstein) statistics for 
nucleons (pions) allows for isospins $I_{NN}$=$I_{\pi\pi}$=0, forbidding 
$I_{NN}$=$I_{\pi\pi}$=1, with weights 2/3 and 1/3, respectively, obtained 
by recoupling the two $P_{33}$ isospins $I_{N\pi}=\frac{3}{2}$ in the $I$=0 
$\Delta\Delta$ state \cite{gg13}. In the general case, for given values of 
$I$, $I_{NN}$ and $I_{\pi\pi}$, we compute the weight $x_I(I_{NN},I_{\pi\pi})$ 
with which ${\vec I}_{NN}+{\vec I}_{\pi\pi}={\vec I}$ is obtained by 
recoupling from ${\vec I}_{N\pi}+{\vec I}_{N\pi}={\vec I}$. 
This is accomplished using a $9j$ recoupling coefficient, 
\begin{equation} 
x_I(I_{NN},I_{\pi\pi})=(2I_{NN}+1)(2I_{\pi\pi}+1)(2I_{N\pi}+1)^2 
\left\{ \begin{matrix} 1/2 & 1 & I_{N\pi} \cr 1/2 & 1 & I_{N\pi} \cr 
I_{NN} & I_{\pi\pi} & I \cr \end{matrix} \right\} ^2,  
\label{eq:9j} 
\end{equation} 
with a similar expression in spin space for $x_J(S_{NN},L_{\pi\pi})$. 
A width-suppression fraction $x_{IJ}$ is defined by summing up over all 
quantum-statistically allowed products: 
\begin{equation} 
x_{IJ}= \sum_{I_{NN},I_{\pi\pi},S_{NN},L_{\pi\pi}} x_I(I_{NN},I_{\pi\pi}) 
x_J(S_{NN},L_{\pi\pi}). 
\label{eq:xIJ} 
\end{equation} 
If the quantum-statistics requirement is relaxed, and summation is extended 
over all possible couplings, then $x_{IJ}=1$ by completeness. The values of 
$x_{IJ}$ according to Eq.~(\ref{eq:xIJ}) are listed in Table~\ref{tab:xIJ}. 

\begin{table}[hbt] 
\begin{center} 
\caption{Values of width-suppression factors $x_{IJ}$ (\ref{eq:xIJ}) for 
$\Delta\Delta$ dibaryons.} 
\begin{tabular}{ccccccccc} 
\hline 
$IJ$ & $01$ & $10$ & $03$ & $30$ & $12$ & $21$ & $23$ & $32$ \\ 
\hline 
$x_{IJ}$ & 13/27 & 13/27 & 2/3 & 2/3 & 14/27 & 14/27 & 1/3 & 1/3 \\ 
\hline 
\end{tabular} 
\label{tab:xIJ} 
\end{center} 
\end{table} 

\subsection{Two-body interactions} 

The $P_{33}$ $\pi N$ interaction was already specified in 
Eqs.~(\ref{eq6})-(\ref{eq7p}), so we need only to construct the $NB$ 
interactions that generate the ${\cal D}_{12}$ and ${\cal D}_{21}$ 
dibaryon resonances. Starting with ${\cal D}_{12}$, we wish to construct 
a separable-potential model that describes the $NN$ $^1D_2$ partial wave 
below and above the $\pi NN$ threshold. The simplest choice would be to 
consider a model that couples the $NN$ and $N\Delta$ two-body channels. 
However, this model will not generate the inelastic $\pi NN$ cut at its 
correct position, since the mass of the $\Delta$ is much higher than 
$m_N+m_\pi$. Therefore we added another $s$-wave $NN'$ channel, where 
$N'$ is an auxiliary stable baryon with quantum numbers $I(J^P)=\frac{1}{2}
(\frac{3}{2}^+)$ and mass $m_{N'}=m_N+m_\pi$. Note that $J_{N'}=\frac{3}{2}$ 
is mandatory in order to connect to $J({\cal D}_{12})=2$, and $I_{N'}=\frac
{1}{2}$ comes natural because the other option $I_{N'}=\frac{3}{2}$ is already 
taken up by the $P_{33}$ channel for the $\pi-N$ isobar $\Delta$ resonance. 
Note also that $N'$, with $\frac{1}{2}(\frac{3}{2}^+)$, has nothing to do with 
the $P_{13}$ $\pi N$ channel. Having introduced the auxiliary $N'$ baryon, 
we fitted the $NN$ amplitude of Arndt et al. \cite{arndt07} in the $^1D_2$ 
partial wave using the three-channel separable potential 
\begin{equation} 
V_1^{mn}(p_1,p'_1)= \lambda_1 g_1^m(p_1) g_1^n(p'_1);\,\,\,\,\,\,\, 
m,n=1-3, 
\label{eq1} 
\end{equation} 
where the three channels are $1=NN$ ($d$-wave), $2=NN'$ and $3=N\Delta$, both 
$s$-wave, with a stable $\Delta$ of mass $m_{\Delta} = 1232$~MeV and quantum 
numbers $I(J^P)=\frac{3}{2}(\frac{3}{2}^+)$. This coupled-channel system 
is written generically as $NB$, where $B$ stands for ($N,N',\Delta$), 
and its $t$-matrix is obtained by solving the Lippmann-Schwinger equation 
with relativistic kinematics,  
\begin{eqnarray} 
t_1^{mn}(\omega_1;p_1,p_1^\prime)&=& V_1^{mn}(p_1,p_1^\prime)+\sum_{r=1}^3 
\int_0^\infty {p_1^{\prime\prime}}^2 dp_1^{\prime\prime} \, 
V_1^{mr}(p_1,p_1^{\prime\prime}) 
\nonumber \\ && \times 
\frac{1}{\omega_1-E_N(p_1^{\prime\prime})-E_r(p_1^{\prime\prime})+i\epsilon} 
t_1^{rn}(\omega_1;p_1^{\prime\prime},p_1^\prime), 
\label{eq2} 
\end{eqnarray} 
which in the case of the separable potential (\ref{eq1}) has the solution 
\begin{equation} 
t_1^{mn}(\omega_1;p_1,p_1')= g_1^m(p_1)\tau_1(\omega_1)g_1^n(p_1^\prime), 
\label{eq3} 
\end{equation} 
where the propagator of the ${\cal D}_{12}$-isobar is expressed through its 
inverse by 
\begin{equation} 
\tau_1^{-1}(\omega_1)=\lambda_1^{-1}-\sum_{r=1}^3 \int_0^\infty p_1^2 dp_1 
\frac{[g_1^r(p_1)]^2} 
{\omega_1-E_N(p_1)-E_r(p_1)+i\epsilon}, 
\label{eq4} 
\end{equation} 
with $m_r=(m_N,m_{N'},m_{\Delta})$ for $r=(1,2,3)$. 
The $r=2$ $NN'$ channel is responsible for generating 
the inelastic cut starting at the $\pi NN$ threshold. 

\begin{figure}[!ht] 
\begin{center} 
\includegraphics[width=0.48\textwidth]{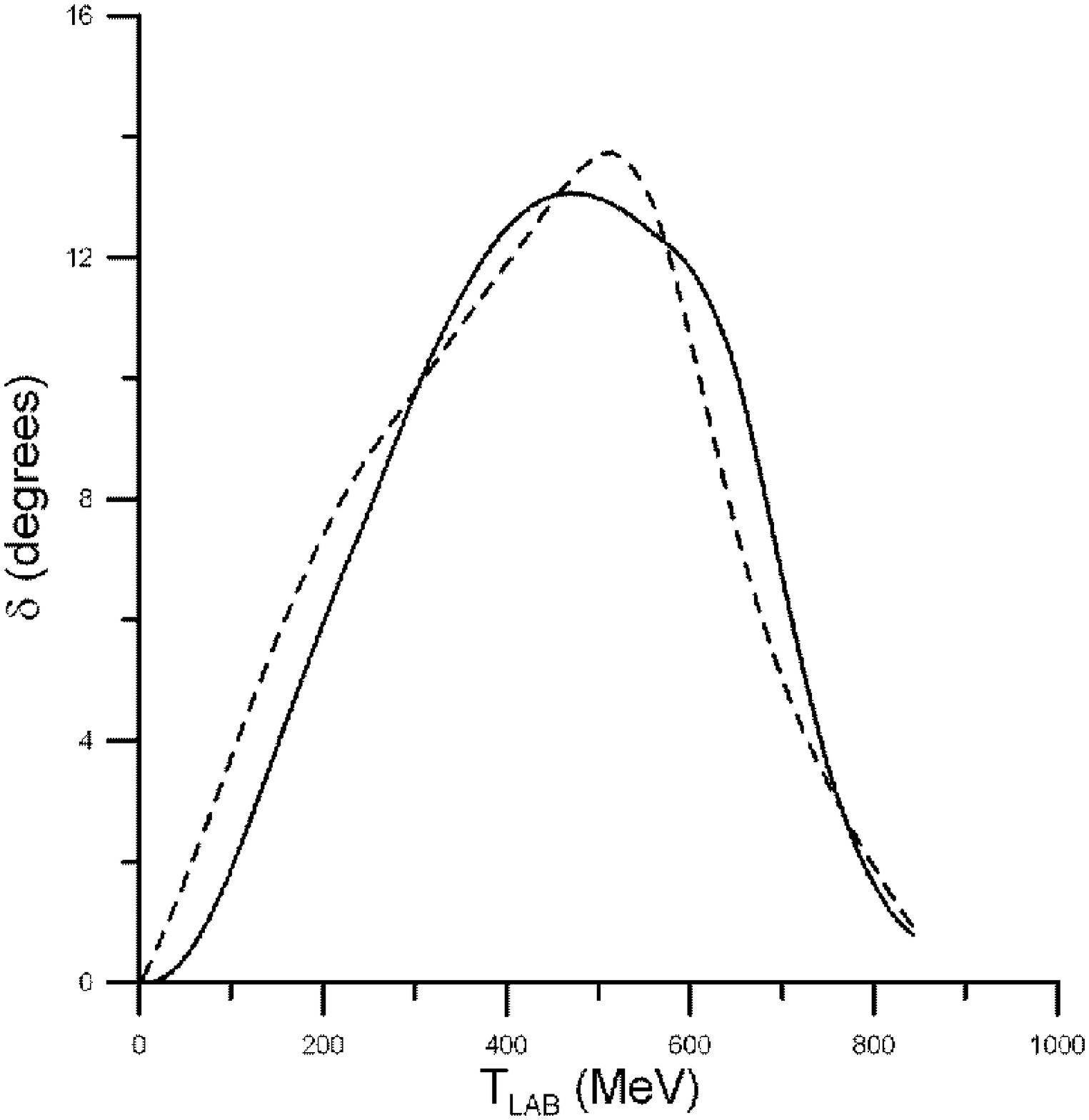} 
\includegraphics[width=0.48\textwidth]{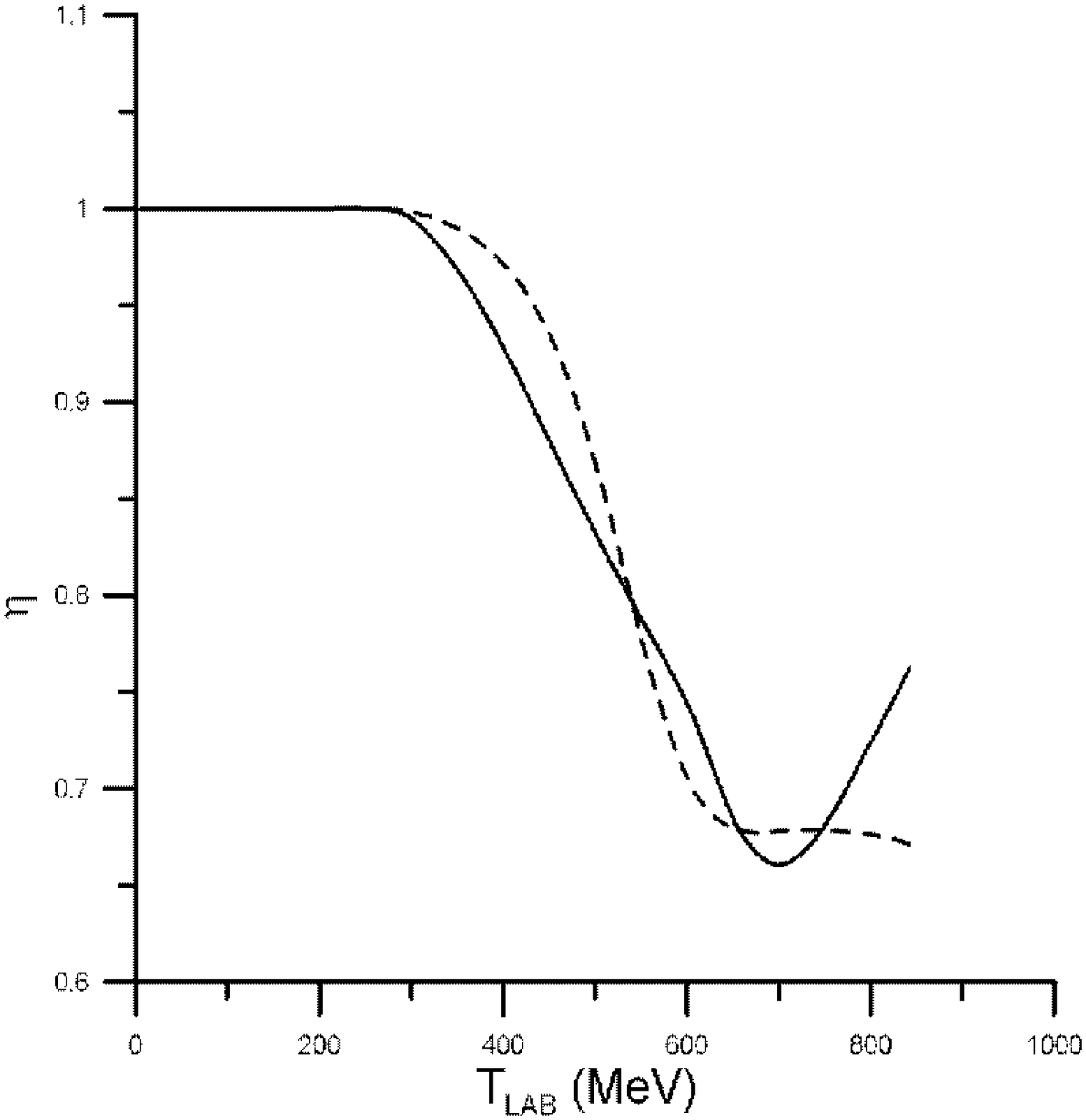} 
\caption{Fits (solid curves) to $NN$ $^1D_2$ scattering parameters 
(dashed curves) $\delta$ (left) and $\eta$ (right) \cite{arndt07}, 
using the $A_1^j=1$ ($j=1,2,3$) type-I best-fit parameters from 
Table~\ref{tab:g1}.} 
\label{fig:1D2} 
\end{center} 
\end{figure}

The form factors of the separable potential (\ref{eq1}) 
were taken in the form (which is termed type I) 
\begin{equation} 
g_1^n(p_1)=\frac{(p_1/o)^\ell}{[1+p_1^2/(\alpha_1^n)^2]^{1+\ell/2}} 
\left[1+A_1^n\frac{(p_1/o)^2} {1+p_1^2/(\alpha_1^n)^2} \right], 
\label{eq5} 
\end{equation} 
where $o=1$ fm$^{-1}$ ensures that the form factors $g_1^n$ have no units, and 
with $\ell=2$ for $n=1$ and $\ell=0$ for $n=$ 2 and 3. These form factors fall 
off as $p_1^{-2}$ upon $p_1\to\infty$. We also considered form factors of 
a form termed type II: 
\begin{equation} 
g_1^n(p_1)=\frac{(p_1/o)^\ell}{[1+p_1^2/(\alpha_1^n)^2]^{3/2+\ell/2}} 
\left[1+A_1^n\frac{(p_1/o)^2} {1+p_1^2/(\alpha_1^n)^2} \right], 
\label{eq5p} 
\end{equation} 
which fall off as $p_1^{-3}$ upon $p_1\to\infty$. The inverse-range parameters 
$\alpha_1^n$ were limited to values $\alpha_1^n \lesssim 3$~fm$^{-1}$ as much 
as possible to ensure that shorter-range degrees of freedom, for example 
$\pi N\to\rho N$, need not explicitly be introduced. Good fits to the $NN$ 
$^1D_2$ scattering parameters satisfying this limitation required that not 
all $A_1^n$ be zero. Best-fit values of $\lambda_1$ and $\alpha_1^n$ in these 
two models were determined by scanning on selected values of $A_1^n$ and are 
listed in Table~\ref{tab:g1}. The fitted $NN$ $^1D_2$ phase shifts $\delta$ 
and inelasticities $\eta$, defined in terms of the $T$-matrix by 
\begin{equation} 
S=1+2{\rm i}T=\eta \exp(2{\rm i}\delta), 
\label{eq5pp} 
\end{equation} 
are compared in Fig.~\ref{fig:1D2} with values derived from $pp$ scattering 
experiments \cite{arndt07}. A variance of 0.02 was used for Re~$T$ and Im~$T$ 
in these fits. We note that the decrease of the inelasticity $\eta$ from 
a value 1 is due to the $r=2$ $NN'$ subchannel which generates the inelastic 
cut starting at the $\pi NN$ threshold, and that no explicit ${\cal D}_{12}$ 
pole term was introduced in the $r=3$ $N\Delta$ subchannel. 
Yet, the three-channel system owns a ${\cal D}_{12}$ pole, listed in the last 
column of Table~\ref{tab:g1}.  

\begin{table}[htb] 
\begin{center} 
\caption{Best-fit parameters $\alpha_1^n$ (fm$^{-1}$) and $\lambda_1$ of the 
three-channel separable potential (\ref{eq1}) with type-I (\ref{eq5}) and 
type-II (\ref{eq5p}) form factors for selected values of $A_1^n$ that provide 
the lowest $\chi^2$, see Fig.~\ref{fig:1D2}, plus pole position $W$ (in MeV) 
of ${\cal D}_{12}$.}
\begin{tabular}{cccccccc} 
\hline 
$g_1$ & $A_1^1$$A_1^2$$A_1^3$ & $\chi^2$/N & $\alpha_1^1$ & $\alpha_1^2$ & 
$\alpha_1^3$ & $\lambda_1$ (fm$^2$) & $W({\cal D}_{12})$ \\ 
\hline 
 I & 0 1 $\frac{3}{2}$ & 1.15 & 2.04 & 2.16 & 2.44 & $-$0.00340 & 2171$-$i45\\ 
II & 0 1 $\frac{3}{2}$ & 1.12 & 2.50 & 2.73 & 3.16 & $-$0.00287 & 2176$-$i49\\ 
 I & 0 1 1 & 1.10 & 2.00 & 2.11 & 2.96 & $-$0.00313 & 2172$-$i58 \\ 
II & 0 1 1 & 1.05 & 2.41 & 2.57 & 3.74 & $-$0.00296 & 2179$-$i61 \\ 
 I & 1 1 1 & 0.78 & 1.47 & 2.27 & 3.24 & $-$0.00214 & 2177$-$i63 \\ 
II & 1 1 1 & 0.70 & 1.81 & 3.17 & 4.44 & $-$0.00132 & 2182$-$i74 \\ 
\hline 
\end{tabular} 
\label{tab:g1} 
\end{center} 
\end{table} 

In the case of the ${\cal D}_{21}$ dibaryon there is no experimental 
information to count on. Since as shown in the previous section 
${\cal D}_{21}$ and ${\cal D}_{12}$ have similar structure and are almost 
degenerate, it is natural to assume that ${\cal D}_{21}$ is generated by 
the same separable potential model that generates ${\cal D}_{12}$. However, 
with isospin $\frac{1}{2}$ constituents, the $NN$ and $NN'$ channels are 
unable to couple to total isospin $I=2$, so alternatively we will replace $N'$ 
by another auxiliary baryon $N''$ with $I(J^P)=\frac{3}{2}(\frac{1}{2}^+)$, 
with the same  fit parameters used for ${\cal D}_{12}$. 

\subsection{$B\Delta - \pi D$ coupled-channel $\pi NB$ Faddeev equations} 

Using standard three-body techniques \cite{gg11} the integral equation 
depicted in Fig.~\ref{fig:piNB} is written explicitly in a vector form, 
generalizing expression (\ref{eq19}) for the $\pi NN$ system: 
\begin{equation} 
T_m^{IJ}(W;q_3)=\sum_{n=1}^3\int_0^\infty dq_3^\prime 
M_{mn}^{IJ}(W;q_3,q_3^\prime){\cal T}_3^n(W;q_3^\prime)T_n^{IJ}(W;q_3^\prime), 
\label{eq10} 
\end{equation} 
where the vectorial indices $m,n=1,2,3$ correspond to the three $D$-isobar 
$NB$ channels ($NN,NN',N\Delta$) or equivalently to the three possible decay 
channels $B\Delta$=($N\Delta,N'\Delta,\Delta\Delta$), and the kernels 
$M_{mn}^{IJ}$ are given by  
\begin{equation} 
M_{mn}^{IJ}(W;q_3,q_3^\prime)=2\sum_{d=1}^2\int_0^\infty dq_1 
K_{31;md}^{IJ}(W;q_3,q_1){\cal T}_{1;d}(W;q_1) 
K_{13;nd}^{IJ}(W;q_1,q_3^\prime), 
\label{eq11} 
\end{equation} 
where $d=1,2$ correspond to the $\pi NN$ isobars $D$ with $IJ=12$ and $IJ=21$. 
The reason for a factor 2 on the r.h.s. of Eq.~(\ref{eq11}) is that in the 
decay of the isobar $D$, $D\to NB$, the nucleon $N$ can originate from each 
one of the constituents of $D$, similarly to the way a factor 2 was justified 
on the r.h.s. of Eq.~(\ref{eq20}). The amplitudes $K_{31;nd}^{IJ}(W;q_3,q_1)$ 
are structured similarly to those specified for the $\pi NN$ system by 
Eq.~(\ref{eq22}):  
\begin{eqnarray} 
K_{31;nd}^{IJ}(W;q_3,q_1)&=& 
K_{13;nd}^{IJ}(W;q_1,q_3)= 
\frac{1}{2}q_3q_1\int_{-1}^1 d{\rm cos}\theta\, 
g_3(p_3)\, g_1^n(p_1) b_{31;nd}^{IJ} 
\nonumber \\ && \times 
\frac{\hat p_3\cdot\hat q_1}{W-E_{\pi}(q_1)-E_N(\vec q_1+\vec q_3)
-E_n(q_3)+i\epsilon}, 
\label{eq12} 
\end{eqnarray} 
where $b_{31;nd}^{IJ}=b_{31}^{IJ}$, as given by Eq.~(\ref{e8c5}), with 
$I_1=J_1=1$, $I_2=J_2=\frac{1}{2}$ and $(I_3,J_3)=(\frac{1}{2},\frac{1}{2}),
(\frac{1}{2},\frac{3}{2}),(\frac{3}{2},\frac{3}{2})$ for $n=(1,2,3)$, 
respectively, $I_{23}=1,J_{23}=2$ for $d=1$ ($D={\cal D}_{12}$) and $I_{23}=2,
J_{23}=1$ for $d=2$ ($D={\cal D}_{21}$). In the latter case only $n=3$ is 
effective, since the channels $n=1,2$ do not couple to ${\cal D}_{21}$. 
Finally, $I_{12}=J_{12}=\frac{3}{2}$ remains as was for $\pi N$.  

The propagators of the $\Delta$ and $D$ isobars are ${\cal T}_3^n(W;q_3)$ 
and ${\cal T}_{1;d}(W;q_1)$, respectively. The expression for 
${\cal T}_3^n(W;q_3)$, for example, is given in the three-body cm frame by 
\begin{equation} 
\left[{\cal T}_3^n(W;q_3)\right]^{-1}=\lambda_3^{-1}-\int_0^\infty p_3^2 dp_3 
\frac{[g_3(p_3)]^2}{W-{\cal E}_3(p_3,q_3)-E_n(q_3)+i\epsilon},  
\label{eq9pp} 
\end{equation} 
slightly generalizing the expression (\ref{eq9ppp}) for $\pi NN$, and 
similarly for ${\cal T}_{1;d}(W;q_1)$. Finally, in the propagator 
(\ref{eq9pp}) for $n=3$, the mass 
of the baryon $B=\Delta$ which up to this point has been 
assumed to be real is modified to include its width by using the $\Delta$ 
pole position \cite{arndt06}, in MeV: 
\begin{equation} 
m_B = 1232 \;\;\; \to \;\;\; W_{\Delta}=1211-{\rm i}x_{IJ}49.5, 
\label{eq8ppp} 
\end{equation}  
where the width-suppression factors $x_{IJ}$ are given in Table~\ref{tab:xIJ}.

\subsection{Results and Discussion} 

The integral equations (\ref{eq10}) were solved for the $\Delta\Delta$ 
dibaryon candidates ${\cal D}_{IJ}$, with (i) $IJ=01,03,23$ 
proceeding exclusively through $D={\cal D}_{12}$ in the $\pi D$ intermediate 
state in Fig.~\ref{fig:piNB}, (ii) $IJ=10,30,32$ proceeding exclusively 
through $D={\cal D}_{21}$, and (iii) $IJ=12,21$ that proceed through both 
choices of $D$. We start by listing results in Table~\ref{tab:D03} for 
${\cal D}_{03}$ because of its apparent relevance to the resonance observed 
recently in the WASA@COSY $pn \to d \pi\pi$ measurements \cite{wasa11}. 
Partial results were given in Ref.~\cite{gg13}. 

\begin{table}[htb]
\begin{center}
\caption{${\cal D}_{03}$ pole position (in MeV) found by solving 
Eqs.~(\ref{eq10}) for the three best-fit baryon-baryon interactions labeled 
by their values of $A_1^1,A_1^2,A_1^3$, in decreasing order of $\chi^2$, 
using combinations of form factors $(g_3^{\rm k},g_1^{\rm k'})$ with k,k' each 
running on types I and II. The half-width values in parentheses disregard 
quantum-statistics correlations, i.e. $x_{03}=1$.} 
\begin{tabular}{ccccc}
\hline
$A_1^1,A_1^2,A_1^3$ & $g_3^{\rm I}$ $g_1^{\rm I}$ & 
$g_3^{\rm I}$ $g_1^{\rm II}$ & $g_3^{\rm II}$ $g_1^{\rm I}$ & 
$g_3^{\rm II}$ $g_1^{\rm II}$ \\
\hline 
0,1,$\frac{3}{2}$ & 2392$-$i52 & 2386$-$i47 & 2380$-$i45 & 2373$-$i41 \\ 
0,1,1 & 2384$-$i44 & 2374$-$i38 & 2356$-$i30 & 2344$-$i26 \\ 
1,1,1 & 2383$-$i41(47) & 2386$-$i38(44) & 2343$-$i24(31) & 2337$-$i21(28) \\ 
\hline 
\end{tabular} 
\label{tab:D03} 
\end{center} 
\end{table} 

The ${\cal D}_{03}$ pole positions listed in Table~\ref{tab:D03} result from  
calculations that use all four combinations of form factors $g_3$ and $g_1$ 
within each of the three lowest $\chi^2$ fits of $V_1$ to the $^1D_2$ $NN$ 
scattering parameters marked by their values of the parameters $A_1^j$ 
($j=1,2,3$) from Table~\ref{tab:g1}. The calculated pole positions are 
sensitive primarily to the choice of $\pi N$ form factor $g_3$ from 
Table~\ref{tab:piN}; the smaller its spatial extension $r_0$, the lower 
the calculated mass values are. 
Admitting values of $r_0$ appreciably below 0.9~fm, the smaller of the two 
values chosen here, calls for the introduction of explicit vector-meson 
and/or quark-gluon degrees of freedom which are outside the scope of the 
present model. The dependence of the calculated pole positions on the chosen 
baryon-baryon form factor $g_1$ of Eqs.~(\ref{eq5}) and (\ref{eq5p}) is 
weaker. For a given choice of $g_1$, the calculated mass values display 
sensitivity primarily through the fitted values of the inverse-range 
parameters $\alpha_1^n$ listed in Table~\ref{tab:g1}, particularly 
$\alpha_1^3$. Whereas values of $\alpha_1^3 \lesssim 2.5$~fm$^{-1}$ 
were found impossible to get, going beyond $\alpha_1^3\sim 3$~fm$^{-1}$ 
was considered undesirable, again requiring the introduction of explicit 
short-range degrees of freedom. For these reasons, the discussion 
below is limited to results obtained using type I form factor $g_1$, 
displaying only the sensitivity to the $\pi N$ form factor $g_3$. 
As for width values $-2{\rm Im}W$, the calculated widths display 
little sensitivity to these form factors and the widths are determined 
primarily by the phase space available for decay. The listed half-widths 
values were calculated using the width-suppression fraction 
$x_{03}=\frac{2}{3}$ from Table~\ref{tab:xIJ}. For comparison we added 
in parentheses for the lowest $\chi^2$ best fit, last line in the table, 
the half-width calculated disregarding quantum-statistics correlations, 
i.e. $x_{03}=1$. The masses are insensitive to the value of $x_{03}$ used 
in the calculations. We conclude this discussion of the calculated 
${\cal D}_{03}$ results by noting that the average over the four results 
shown in the table for the best fit potential (last line) comes very 
close to the reported mass $M=2.37$~GeV and width $\Gamma\approx$~70~MeV 
of the ${\cal D}_{03}$ resonance~\cite{wasa11}. 

\begin{table}[htb]
\begin{center}
\caption{${\cal D}_{03}$ and ${\cal D}_{30}$ pole positions (in MeV) found 
by solving Eqs.~(\ref{eq10}) with $\pi N$ form factor $g_3^{\rm k}$ of types 
k=I,II. $N'$ denotes the best-fit $NN-NN'-N\Delta$ coupled channel interaction 
$V_1$ (last line in Table~\ref{tab:D03} for ${\cal D}_{03}$). $N''$ stands 
for replacing $N'(I=\frac{1}{2},J=\frac{3}{2})$ by $N''(I=\frac{3}{2},J=\frac
{1}{2})$ in the ${\cal D}_{30}$ calculation, retaining form factors.} 
\begin{tabular}{ccccc} 
\hline 
${\cal D}_{IJ}$ & $g_3^{\rm I}$ $N'$ & $g_3^{\rm I}$ $N''$ & 
$g_3^{\rm II}$ $N'$ & $g_3^{\rm II}$ $N''$ \\ 
\hline 
${\cal D}_{03}$ & 2383$-{\rm i}$41(47) & 2383$-{\rm i}$41(47) & 
2343$-{\rm i}$24(31) & 2343$-{\rm i}$24(31) \\ 
${\cal D}_{30}$ & 2411$-{\rm i}$41(49) & 2391$-{\rm i}$39(46) & 
2370$-{\rm i}$22(30) & 2350$-{\rm i}$22(29) \\
\hline 
\end{tabular} 
\label{tab:D30} 
\end{center} 
\end{table} 

We proceed now to discuss the exotic $\Delta\Delta$ dibaryon candidate 
${\cal D}_{30}$, noticing that since in Eq.~(\ref{eq12}) $b_{31;31}^{03}
=b_{31;32}^{30}$ the states $IJ=03$ and $IJ=30$ become degenerate in the 
limit of equal $D={\cal D}_{12}$ and $D={\cal D}_{21}$ isobar propagators. 
Since $D={\cal D}_{12}$ was found to lie lower than $D={\cal D}_{21}$, we 
expect also ${\cal D}_{03}$ to lie lower than ${\cal D}_{30}$. The results 
of our Faddeev calculations, presented in Table~\ref{tab:D30}, indeed confirm 
this expectation, placing ${\cal D}_{30}$ 28~MeV above ${\cal D}_{03}$ for 
the standard calculation marked $N'$ in the table, and only 8~MeV apart 
for the calculation in which $N'$ was replaced by $N''$ (with $I(J^P)=\frac
{3}{2}(\frac{1}{2}^+)$ and same fit parameters as used for ${\cal D}_{12}$). 
Such approximate degeneracy was noticed in old OBEP work \cite{kamae77} and 
in several of the quark-based works \cite{oka80,malt85,valc01,mota02,huang13}, 
and it has been discussed recently in Ref.~\cite{BBC13}. In our case it 
is just a consequence of the approximate $I \leftrightarrow J$ underlying 
symmetry of our model. 

\begin{table}[htb]
\begin{center}
\caption{${\cal D}_{IJ}$ pole positions (in MeV) found by solving
Eqs.~(\ref{eq10}) for the best-fit baryon-baryon interaction $V_1$ with type I 
$g_1$ form factors (\ref{eq5}) and $A_1^j=1, j=1,2,3$, using width-suppression 
fractions $x_{IJ}$ from Table~\ref{tab:xIJ} and type I,II $g_3$ $\pi N$ form 
factors, with averaged results denoted $W$ in the last line. 
Note: ${\cal D}_{23}$ is numerically unstable.} 
\begin{tabular}{ccccccc} 
\hline 
$g_3$ & ${\cal D}_{03}$ & ${\cal D}_{30}$ & ${\cal D}_{12}^\ast$ & 
${\cal D}_{21}^\ast$ & ${\cal D}_{23}$ & ${\cal D}_{32}$ \\
\hline 
I & 2383$-$i41 & 2411$-$i41 & 2431$-$i76 & 2449$-$i94 & 2431$-$i72 & 
2444$-$i89 \\ 
II & 2343$-$i24 & 2370$-$i22 & 2428$-$i67 & 2436$-$i72 & 2429$-$i72 & 
2439$-$i66 \\ 
$W$ & 2363$-$i33 & 2391$-$i32 & 2430$-$i72 & 2443$-$i83 & 2430$-$i72 & 
2442$-$i78 \\ 
\hline 
\end{tabular} 
\label{tab:DIJ} 
\end{center} 
\end{table} 

The ${\cal D}_{03}$ and ${\cal D}_{30}$ are not the only $\Delta\Delta$ 
dibaryon candidates found as resonances in our Faddeev calculations. In 
Table~\ref{tab:DIJ} we list all the ${\cal D}_{IJ}$ resonance poles found 
using the best-fit $V_1$ for two choices of the $\pi N$ form factor $g_3$. 
Averaged results are also listed. The table suggests that in addition to 
the (${\cal D}_{03},{\cal D}_{30}$) doublet, the lowest of all $\Delta\Delta$ 
dibaryon doublets, two additional $I\leftrightarrow J$ doublets are found 
several tens of MeV higher in energy and are twice or more as broad: 
(${\cal D}_{12}^\ast,{\cal D}_{21}^\ast$), where the asterisk distinguishes 
these excited $IJ=$ $12$ and $21$ resonances from the lower (${\cal D}_{12},
{\cal D}_{21}$) $N\Delta$ dibaryon resonances of Sect.~\ref{sec:NDel}, 
and the (${\cal D}_{23},{\cal D}_{32}$) doublet of resonances. 
To understand this hierarchy we recall that by Eq.~(\ref{eq12}) 
the kernel $M^{IJ}$ (\ref{eq11}) is roughly proportional to the squares 
of the recoupling coefficients $b_{31;nd}^{IJ}$. For ${\cal D}_{03}$ 
and ${\cal D}_{30}$ the squares of the only nonvanishing coefficients 
$b_{31;31}^{03}$ and $b_{31;32}^{30}$, respectively, assume the maximal 
value 1. For the other two doublets, several nonvanishing coefficients 
contribute with average square about 0.5, so that the effective interactions 
in these systems are weaker than for ${\cal D}_{03}$ and ${\cal D}_{30}$. 
It is interesting to note that for the nonresonant dibaryon candidates 
${\cal D}_{01}^\ast$ and ${\cal D}_{10}^\ast$ the squares of their only 
nonvanishing coefficients $b_{31;31}^{01}$ and $b_{31;32}^{10}$, respectively, 
are 0.125 each, exceedingly small to form resonances. 


In order to understand the mechanism for the relatively small widths of 
${\cal D}_{03}$ and ${\cal D}_{30}$, recall the formulation of the three-body 
$\pi NB$ model and its $B\Delta - \pi D$ coupled-channel description around 
Eq.~(\ref{eq:D}), where $D$ corresponds to a three-channel $NB\equiv 
(NN,NN',N\Delta)$ isobar. The decay of ${\cal D}_{03}$ and ${\cal D}_{30}$ 
through the upper $B\Delta \equiv (N\Delta,N'\Delta,\Delta\Delta)$ channel 
can proceed only via the $\Delta\Delta$ subchannel. In particular, the strong 
decay expected from the $N'\Delta$ subchannel is forbidden and the resulting 
effective decay width is reduced to $\approx$(0.3--0.5)$\Gamma_{\Delta}$ in 
terms of the free-space $\Delta$ decay width $\Gamma_{\Delta}\approx 120$~MeV. 
Similar suppression should have occurred for the decay width of 
${\cal D}_{32}$, but its higher mass provides larger phase space for decay, 
also elastically to the $\pi D$ channel. Finally, the ${\cal D}_{12}^\ast$, 
${\cal D}_{21}^\ast$ and ${\cal D}_{23}$ dibaryon resonances are allowed to 
decay through all of the $B\Delta$ subchannels and their decay width is not 
suppressed with respect to $\Gamma_{\Delta}$. 

\section{Summary and Outlook} 
\label{sec:concl}

A unified hadronic approach to the calculation of non-strange dibaryon 
candidates was presented in this work. The building blocks of the model here 
applied are nucleons, $\Delta$'s and pions, the latter playing a special role. 
Apart from generating long-range pion-exchange interactions, as in the first 
diagram on the r.h.s. of the LS equation Fig.~\ref{fig:piNN}, the pion forms 
a $\Delta$ resonance by scattering off a nucleon, thereby linking the two 
baryons of the model. A $\pi NN$ three-body model was formulated in terms of 
Faddeev equations to explore $N\Delta$ dibaryons. Separable interactions were 
fitted to scattering phase shifts in the dominant $NN$ $s$-wave channels and 
the $\pi N$ $P_{33}$ channel. With this input, the $\pi NN$ Faddeev equations 
were solved using relativistic kinematics. Resonance poles in the 
$I(J^P)=1(2^+),2(1^+)$ $N\Delta$ channels were found nominally below 
threshold and were attributed to the $N\Delta$ dibaryon candidates 
${\cal D}_{12},{\cal D}_{21}$ predicted by Dyson and Xuong \cite{dyson64}. 
The calculated $I(J^P)=1(2^+)$ resonance mass and width agree closely with 
those extracted phenomenologically from $NN$ and $\pi d$ scattering and 
reaction data \cite{igor84,arndt87,hosh92}. The existence of the ``exotic" 
$I(J^P)=2(1^+)$ resonance, in contrast, lacks experimental support or 
phenomenological evidence because with isospin $I=2$ it is decoupled from $NN$ 
scattering data. Of course, given the proximity of these nearly degenerate 
$N\Delta$ resonances to the $N\Delta$ threshold, and given that their widths 
are similar to that of a free $\Delta$, it is not an easy task to distinguish 
them from $N\Delta$ threshold effects. 

To study $\Delta\Delta$ dibaryons we formulated a $\pi NB$ three-body 
model with pairwise separable interactions in the dominant $\pi N$ $P_{33}$ 
channel, as above, and in the $NB$ dibaryon $I(J^P)=1(2^+),2(1^+)$ resonating 
channels. The $I(J^P)=1(2^+)$ interaction was constrained by the $NN$ $^1D_2$ 
scattering data without explicitly assuming it to resonate. Special care was 
taken to ensure that the inverse-range parameters $\alpha_1^n$ of the $NB$ 
interaction satisfy the constraint $\alpha_1^n\lesssim 3$~fm$^{-1}$ to be 
consistent with the exclusion of explicit vector mesons and shorter-range 
degrees of freedom from our long-range physics model. The $I(J^P)=1(2^+)$ 
interaction was also employed in the $2(1^+)$ channel in most of the reported 
calculations. With these input two-body interactions, the $\pi NB$ Faddeev 
equations were solved, allowing the $\Delta$ constituent of the model to 
acquire decay width compatible with the requirements of quantum statistics 
with respect to the pion and nucleon constituents of the model. Several 
${\cal D}_{IJ}$ dibaryon resonances were found below the $\Delta\Delta$ 
threshold, notably the (${\cal D}_{03},{\cal D}_{30}$) doublet, with 
${\cal D}_{03}$ the lowest dibaryon at complex energy value 2363$-$i33~MeV, 
where the theoretical uncertainty of its mass and width values is estimated 
by $\pm$20~MeV, in good agreement with the resonance observed by WASA@COSY 
in double-pion production $pn\to d\pi\pi$ reactions \cite{wasa11}. 

It is remarkable that our long-range physics model calculations reproduce 
the two nonstrange dibaryons established experimentally and phenomenologically 
so far, the $N\Delta$ dibaryon ${\cal D}_{12}$ \cite{igor84,arndt87,hosh92} 
and the $\Delta\Delta$ dibaryon ${\cal D}_{03}$ reported by the WASA@COSY 
Collaboration \cite{wasa11}. Among the other dibaryon candidates predicted 
to resonate in our model calculations, the broad ${\cal D}_{12}^\ast(2430)$ 
($\Gamma\approx$140~MeV) deserves attention. It would be useful to place 
constraints on the appearance of this dibaryon candidate in partial-wave 
analyses of the $NN$ $^1D_2$ wave. The other predicted dibaryons, 
a relatively narrow ${\cal D}_{30}(2390)$ ($\Gamma\approx$65~MeV) 
and a doublet of broad resonances (${\cal D}_{23},{\cal D}_{32}$) at 2440 MeV 
($\Gamma\approx$160--170~MeV) are all ``exotic" in the sense that their high 
value of isospin forbids them to couple to $NN$ partial waves. Among these 
``exotic" dibaryon candidates, ${\cal D}_{30}(2390)$ is particularly 
interesting. It was highlighted recently by Bashkanov, Brodsky and Clement 
\cite{BBC13} who focused attention to the special but unspecified role played 
by six-quark hidden-color configurations in forming the (${\cal D}_{03},
{\cal D}_{30}$) dibaryon resonances. However, the recent quark-based 
calculations by Huang, Ping and Wang \cite{huang13} conclude that such 
configurations enhance binding by merely 15$\pm$5~MeV, which is within 
the theoretical uncertainty claimed in our hadronic-basis calculations.

\section*{Acknowledgments} The research of A.G. is supported partially by the 
HadronPhysics3 networks SPHERE and LEANNIS of the European FP7 initiative. 
H.G. is supported in part by COFAA-IPN (M\'exico).

\end{document}